\documentclass{aa}
\def\he{He~{\sc{i}}\ }
\def\hee{He~{\sc{ii}}\ }
\def\nn{N~{\sc{iii}}\ }
\def\n2{N~{\sc{ii}}\ }

\def\si{Si~{\sc{iv}}\ }
\def\ss{Si~{\sc{iii}}\ }
\def\cc{C~{\sc{iii}}\ }
\def\cd{C~{\sc{ii}}\ }
\def\o3{O~{\sc{iii}}\ }

\def\od{O~{\sc{ii}}\ }
\def\hd{HD\,108}
\def\ha{H$\alpha$}
\def\hb{H$\beta$}
\def\hg{H$\gamma$}
\usepackage{graphicx}
\usepackage{times}
\begin{document}
\title{What is the real nature of \hd\ ? \thanks{Based on observations collected at the Observatoire de Haute-Provence (France).}}
\author{Y. Naz\'e\thanks{Research Fellow FNRS (Belgium)} \and J.-M. Vreux \and G. Rauw\thanks{Postdoctoral Researcher FNRS (Belgium)}} 
\offprints{Y. Naz\'e}
\mail{naze@astro.ulg.ac.be}
\institute{Institut d'Astrophysique et de G\'eophysique, Universit\'e de Li\`ege,  5, Avenue de Cointe, B-4000 Li\`ege, Belgium}

\abstract{Since the beginning of the past century, the nature of \hd\ has been a subject of intense debate. One after another, astronomers explored its variability and attributed it either to binarity, or to changes in the stellar wind of a single star. In this article, we analyse a 30 years long campaign of spectroscopic observations of this star with special emphasis on the last 15 years during which photographic plates have been replaced by CCD detectors. Our investigation of the radial velocities of \hd\ yields no significant short- or long-term period and does not confirm the published periodicities either. Though the radial velocity of \hd\ appears clearly variable, the variations cannot be explained by the orbital motion in a spectroscopic binary. However, our data reveal spectacular changes in the H~{\sc{i}} Balmer lines and some \he profiles over the years. These lines continuously evolved from P Cygni profiles to `pure' absorption lines. A similar behaviour was already observed in the past, suggesting that these changes are recurrent. \hd\ seems to share several characteristics of Oe stars and we discuss different hypotheses for the origin of the observed long-term variations. As we are now in a transition period, a continuous monitoring of \hd\ should be considered for the next few years. 
\keywords{stars: early-type -- stars: mass-loss --
          stars: individual: HD\,108}
}
\maketitle

\section{Introduction}
\hd\ has stimulated a lot of controversy over the past 75 years. One of the first detailed studies of its spectrum was presented by Andrillat et al.\ (\cite{and}), who also reviewed the line profile changes (from P Cygni to pure emission/absorption lines). Some years later, using the radial velocities of the \hee $\lambda\lambda$\,4200, 4542 absorption lines, Hutchings (\cite{hut1}) derived an orbital period of 4.612 days, but this result was not confirmed by Vreux \& Conti (\cite{vr}). Vreux \& Conti (\cite{vr}) suggested that the radial velocity variations were rather due to wind variability and though they pointed out the possibility of a 1.02 day period, they concluded that the variability of \hd\ was most probably not periodic. On the basis of Hutchings's work, Bekenstein (\cite{bek}) suggested that \hd\ had undergone a type II supernova without breakup of the binary. Combining radial velocity data from earlier studies with their own measurements, Aslanov \& Barannikov (\cite{asl}) argued in favor of a 5.7937 days period, again attributed to binarity. But the debate about the real nature of \hd\ was not settled. More recently, Underhill (\cite{und2}) found no periodic variation of the radial velocities, thus rejecting the binary hypothesis. She suggested instead that \hd\ was a single star surrounded by a disc and displaying jets that were almost perpendicular to the line of sight. According to Underhill (\cite{und2}), this particular geometry accounts for the peculiar line profile morphology and could explain the spectral variations seen in \hd. Using V-band polarimetric observations, Fox \& Hines (\cite{fox}) did not find any evidence of a binary nature of \hd, but their results did not allow them to choose between large-scale jets or small-scale blobs as the structures triggering the variability. On the contrary, Barannikov (\cite{bar}) combining his own recent radial velocity measurements with all older ones, claimed the existence of a longer period of 1627.6 days and concluded again that the star is indeed a binary.\\
Other properties of \hd\ are also intriguing : its UV spectrum suggests a low mass-loss rate (Hutchings \& van Heteren \cite{hut2}), incompatible with the one derived from its optical and IR (Ferrari-Toniolo et al.\ \cite{fer}) spectrum. Moreover, some authors marked \hd\ as a runaway star, though they did not agree on the exact value of its peculiar velocity. \\

In this paper, we will analyse CCD spectra collected over fifteen years. The observations are presented in Section 2, and we investigate the main spectral characteristics of \hd\ in Section 3. In Section 4, we will discuss the short- and long-term variations of the radial velocities and search for a periodic behaviour of \hd. In the next sections, we analyse the variability of the equivalent widths and the line profiles. Finally, we examine the photometric data and the proper motion of \hd\ before we conclude in Section 8.

\section{Observations and Data Reduction}

\begin{table}[htb]
\caption { \label{tab obs} Summary of our CCD spectroscopic observations. $N$ indicates the total number of spectra obtained during a run.} 
\begin{tabular} {l r r r r r}
\hline
 Date & Telesc. & Wav. range & $N$ & S/N & Disp.\\
& & & & &(\AA\,mm$^{-1}$)\\
\hline\hline
July 1986 & 1.93m&3875-4320& 2& 275&33\\
July 1987& 1.93m&3940-4400& 20& 280&33\\
Aug. 1987&1.93m& 4415-4915& 1& 200&33\\
Aug. 1989& 1.93m&3875-4915& 1& 300& 130\\
Sep. 1989& 1.93m&8430-10400& 1 & 25&260\\
 & 1.93m&8440-11130 & 1& 290& 260\\
Aug. 1990& 1.93m&8450-8765& 1&240 &33 \\
& 1.93m&6830-7130& 1&150&  33\\
& 1.93m&6840-9945& 2&600& 260\\
Aug. 1991& 1.93m&4250-4660& 6&360 & 33\\
Oct. 1992& 1.93m&8450-8765& 3&200 & 33\\
Oct. 1993& 1.93m&3940-4360& 2&220 & 33\\
Aug. 1994& 1.93m&3930-4360& 7&530 & 33\\
Aug. 1996& 1.52m&4065-4925& 3&230 & 33\\
Feb. 1997& 1.52m&6500-6710& 4&150 & 8\\
July 1997& 1.52m&4100-4960& 3&280 & 33\\
Sep. 1998& 1.52m&4455-4885& 3&400 & 16\\
Nov. 1998& 1.52m&4435-4560&15&100 & 5\\
July 1999& 1.52m&4060-4930& 3&300 & 33\\
Aug. 1999& 1.52m&4060-4930& 4&300 & 33\\
Sep. 2000& 1.52m&4455-4905& 11&300& 16\\
\hline
\end{tabular}
\end {table}

An extensive spectroscopic survey of \hd\ was conducted from 1986 to 2000 with the Carelec and the Aur\'elie spectrographs, fed respectively by the 1.93\,m and 1.52\,m telescopes of the Observatoire de Haute-Provence. For Carelec, the detector used was a thin back illuminated RCA CCD with 323$\times$512 pixels of 30$\mu$m squared. For Aur\'elie, it was a TH7832 linear array with a pixel size of 13$\mu$m until 1999. In 2000, this detector was replaced by a 2048$\times$1024 CCD EEV 42-20\#3, whose pixel size is 13.5$\mu$m squared. The exact wavelength ranges and mean S/N are given in Table \ref{tab obs}, together with the dispersion. All the data were reduced in the standard way using the IHAP and MIDAS softwares developed at ESO. The spectra were normalized by fitting splines through carefully chosen continuum windows. Some older photographic plates taken in 1971-72, 1974 and 1976 at OHP were also used to study the line profile variability.

\section{The spectrum of \hd}

\begin{figure*}
\resizebox{\hsize}{!}{\includegraphics{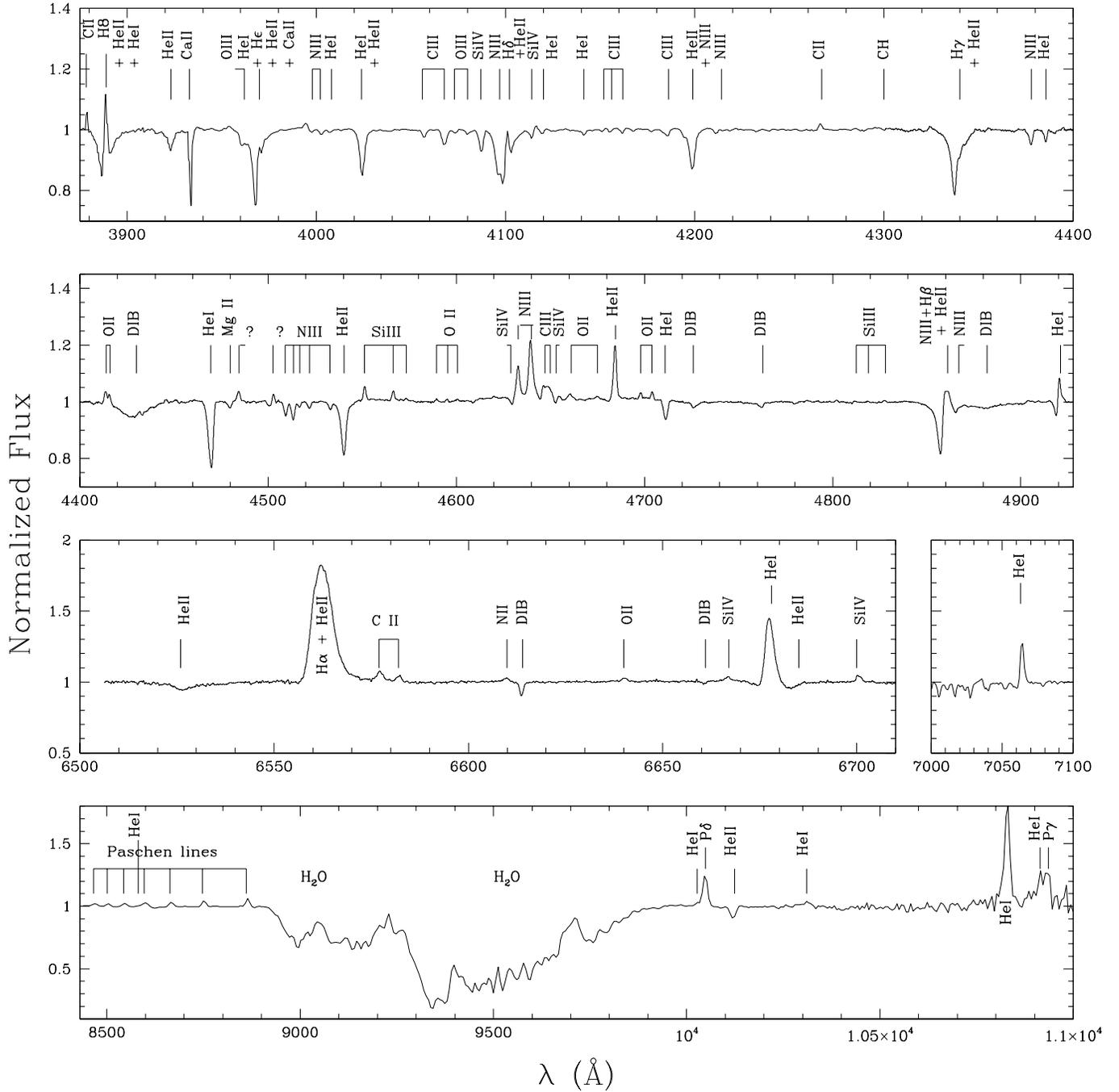}}
\caption{ \label{fig spec} Composite spectrum of \hd\ from blue to IR : the $\lambda\lambda$\,3875-3942 domain comes from the 1986 data, 3942-4300 from the 1994 data, 4300-4930 from the 1999 data, 6500-6710 from the 1997 data, 7000-7100 from the 1990 data and 8430-11000 from the 1989 data.}
\end{figure*}

The spectrum of \hd\ is quite peculiar, showing a great number of emission lines. The blue part of the spectrum is dominated by the presence of strong \nn $\lambda\lambda$\,4634-41 and \hee $\lambda$\,4686 emission lines (see Fig. \ref{fig spec}). As for many Of stars (Underhill et al.\ \cite{und1}), these lines stand on top of a broad emission bump between 4600 and 4700\AA. In addition, \ss $\lambda\lambda$\,4552, 4568, 4574 and various O~{\sc{ii}}\ emission lines, as well as the unidentified Of emissions situated near $\lambda\lambda$\,4485, 4504 are also present. All Balmer lines except \ha\ appear as P Cygni profiles during most of our observing runs, but their morphology changes with time (see Section 6). On the contrary, some lines always appear in absorption, for example \si $\lambda\lambda$\,4088, 4116 and \hee $\lambda\lambda$\,4200, 4542. Some interstellar features can also be seen in Fig \ref{fig spec}: besides the Ca~{\sc{ii}}\ H and K lines, there are diffuse interstellar bands (DIB's) around 4428, 4726 and 4763 \AA, as well as a CH line at 4300 \AA.\\
The H$\alpha$ and \he $\lambda\lambda$\,6678, 7065 emissions dominate the red part of the spectrum. In our 1989 near-IR spectrum, the Paschen lines of Hydrogen appear in emission, as well as \he $\lambda$\,10830 which is very intense. \hee $\lambda$\,10124 is also present, but in absorption.\\

\subsection{Spectral type}
\hd\ has been classified successively as O6\,fqk\footnote{The `f' tag stands for \nn\ $\lambda\lambda$\,4634, 4641 and \hee $\lambda$\,4686 present in emission, the `q' is for the presence of P Cygni profiles and `k' for the presence of Ca lines which are of interstellar nature.} (Plaskett \& Pearce \cite{pla}), O8\,fp (Morgan et al.\ \cite{mor}), O7\,If (Conti \& Alschuler \cite{contal}), O6:f?pe (Walborn \cite{wal}) and finally O7\,IIIfpe (Underhill \cite{und2}). \\
Applying Conti \& Alschuler's (\cite{contal}) criterion to our data, the spectral type varies between O4 (Aug. 1987) and O7.5 (Sept. 2000). A similar type of variations has already been reported by Beals (\cite{bea}), who found for \hd\ an O6 type in 1938 and O7 in 1945. A closer inspection of this situation indicates that the \hee $\lambda$\,4542 line profile is quite constant while the spectral type variations are due to the varying shape of the \he $\lambda$\,4471 line which sometimes appears with a P Cygni profile (see Sections 5 \& 6), revealing that this line is not completely of photospheric origin. This casts doubt on the possibility to apply the Conti \& Alschuler criterion to this star.\\
For the luminosity class, we use the criterion based on the value of $\log(W'') = \log{\frac{EW(\rm{{Si~{\sc{IV}}\ }} \lambda\,4088)}{EW(\rm{{He~{\sc{I}}\ }} \lambda\,4143)}}$. The equivalent widths (EWs) of these two absorption lines show no strong variations, and result in a supergiant classification, regardless of the year of observation. Even though it is possible that \he $\lambda$\,4143 could also be contaminated by a variable emission component, this effect should be small with regard to the stability of our result. Using polarimetric data, Fox \& Hines (\cite{fox}) also favored a supergiant classification. The presence of a strong \he $\lambda$\,10830 emission is also consistent with a supergiant classification, although this emission is also observed in Oe-type objects (Vreux \& Andrillat \cite{vr1}).\\
The presence of \nn $\lambda\lambda$\,4634-41 and \hee $\lambda$\,4686 in emission justifies an `f' tag. Moreover, a `p' tag can be added because of the presence of many emission lines, more numerous than those commonly found in Of spectra. Finally, the emission in H~{\sc{i}} Balmer lines results in the addition of an `e' tag. Choosing the latest spectra, i.e.\ the least affected by possible emission in He~{\sc i}, we can derive a probable O7.5\,Ifpe type for \hd. If the \he $\lambda$\,4471 absorption continues to strengthen over the next years (see Sections 5 \& 6), it is possible that we will finally recover an O8 type as found by Morgan et al.\ (\cite{mor}).

\section{Is \hd\ a spectroscopic binary?}
Using the rest wavelengths listed in Table \ref{tab wavel}, we determined the radial velocities by fitting gaussians to the bottom (top) end of the absorption (emission) lines. The radial velocities (RVs) of the \hee $\lambda\lambda$\,4200, 4542 lines are summarized in Table\,\ref{tab rv}, together with the RVs of the \nn $\lambda\lambda$\,4634,4641 and \hee $\lambda$\,4686 emission lines. We have used the generalized spectrogram technique of Heck et al.\ (\cite{HMM}, hereafter HMM) and the trial method of Lafler \& Kinman (\cite{LK}) to search for periodicities in the \hee $\lambda\lambda$\,4200, 4542 RV time series. For the HMM method, we evaluate a rough estimate of the significance level (SL) by $ SL = e^{-\frac{P}{\sigma^2}}$ where P is the periodogram amplitude and $\sigma$ the rms dispersion of the data. We will consider that a period is significant if its SL is less than 0.01.

\begin{table*}
\caption { \label{tab rv} Mean heliocentric radial velocities of \hee $\lambda\lambda$\,4200, 4542. We have also included the RVs of \nn $\lambda\lambda$\,4634, 4641 and \hee $\lambda$\,4686 for completeness. The date of the observation is given in the format HJD-2440000. } 
\begin{minipage}{9cm}
\begin{tabular} {l r r r r}
\hline
Run & \multicolumn{1}{c}{Date} & RV$_{4200/4542}$ & RV$_{4634/4641}$ & RV$_{4686}$\\
 &  & km\,s$^{-1}$ & km\,s$^{-1}$& km\,s$^{-1}$\\
\hline\hline
Jul.\ 1986& 6627.582& $-66.4$& &\\
& 6628.548& $-47.0$& &\\
Jul.\ 1987& 7007.536& $-68.5$& &\\
& 7007.544& $-79.2$& &\\
& 7007.549& $-72.9$& &\\
& 7007.554& $-74.2$& &\\
& 7008.502& $-84.4$& &\\
& 7008.517& $-80.4$& &\\
& 7008.524& $-84.5$& &\\
& 7009.526& $-82.9$& &\\
& 7009.538& $-78.4$& &\\
& 7010.549& $-74.8$& &\\
& 7010.558& $-73.4$& &\\
& 7011.554& $-72.8$& &\\
& 7011.561& $-75.8$& &\\
& 7011.568& $-77.4$& &\\
& 7011.572& $-74.9$& &\\
& 7012.566& $-77.2$& &\\
& 7012.573& $-71.9$& &\\
& 7013.540& $-70.7$& &\\
& 7013.545& $-64.7$& &\\
& 7013.549& $-70.2$& &\\
Aug.\ 1987& 7016.601& $-69.2$& $-75.4$& $-72.2$\\
Aug.\ 1991& 8490.478& $-93.5$& $-93.3$&\\
& 8490.489& $-86.4$& $-85.1$&\\
& 8490.505& $-85.2$& $-84.0$&\\
& 8492.739& $-94.2$& $-89.9$&\\
& 8496.590& $-84.6$& $-80.4$&\\
& 8497.581& $-77.0$& $-72.6$&\\
Oct.\ 1993& 9264.447& $-61.4$& &\\
& 9264.452& $-67.3$& &\\
Aug.\ 1994& 9576.553& $-44.2$& &\\
& 9577.542& $-64.0$& &\\
& 9579.522& $-64.8$& &\\
& 9580.481& $-58.0$& &\\
& 9581.535& $-58.8$& &\\
& 9582.511& $-63.8$& &\\
& 9583.533& $-59.3$& &\\
Aug.\ 1996& 10316.650& $-58.9$& $-63.4$&$-56.9$\\
& 10316.658& $-58.3$& $-63.0$&$-57.0$\\
\hline
\end{tabular}
\end {minipage}
\begin{minipage}{9cm}
\begin{tabular} {l r r r r}
\hline
Run & \multicolumn{1}{c}{Date}  &RV$_{4200/4542}$ & RV$_{4634/4641}$ & RV$_{4686}$\\
 &  & km\,s$^{-1}$ & km\,s$^{-1}$& km\,s$^{-1}$\\
\hline\hline
Aug.\ 1996 & 10318.644& $-56.8$& $-66.8$&$-53.9$\\
Jul.\ 1997& 10638.574& $-60.8$& $-63.0$&$-41.4$\\
& 10639.557& $-61.7$& $-61.5$&$-54.8$\\
& 10642.573& $-59.1$& $-60.3$&$-53.4$\\
Sep.\ 1998& 11066.635& $-64.1$& $-77.0$&$-65.6$\\
& 11070.629& $-65.0$& $-76.8$&$-64.9$\\
& 11072.638& $-62.8$& $-76.6$&$-61.7$\\
Nov.\ 1998& 11132.377& $-73.1$& \\
& 11132.421& $-66.5$& \\
& 11133.330& $-69.4$& \\
& 11133.380& $-68.2$& \\
& 11133.432& $-69.7$& \\
& 11133.589& $-71.6$& \\
& 11134.378& $-67.9$& \\
& 11134.448& $-66.9$& \\
& 11135.339& $-68.4$& \\
& 11135.394& $-69.8$& \\
& 11135.458& $-66.4$& \\
& 11136.373& $-67.0$& \\
& 11136.439& $-69.5$& \\
& 11137.401& $-67.6$& \\
& 11137.467& $-67.8$& \\
Jul.\ 1999& 11374.597& $-70.1$& $-75.7$&$-68.5$\\
& 11376.532& $-75.7$& $-78.2$&$-69.6$\\
& 11378.530& $-71.2$& $-75.8$&$-65.1$\\
Aug.\ 1999& 11396.600& $-68.8$& $-74.4$&$-66.4$\\
& 11403.635& $-73.4$& $-79.2$&$-68.9$\\
& 11406.634& $-69.7$& $-76.0$&$-67.1$\\
& 11407.639& $-71.6$& $-77.3$&$-69.4$\\
Sep.\ 2000& 11810.586& $-69.4$& $-79.6$&$-66.9$\\
& 11810.595& $-70.6$& $-79.6$&$-67.5$\\
& 11811.567& $-68.6$& $-78.5$&$-67.4$\\
& 11811.575& $-68.6$& $-78.9$&$-67.2$\\
& 11812.594& $-71.6$& $-80.0$&$-70.4$\\
& 11813.603& $-67.8$& $-76.5$&$-67.8$\\
& 11814.595& $-69.7$& $-79.4$&$-66.2$\\
& 11815.596& $-69.8$& $-79.8$&$-69.4$\\
& 11818.531& $-66.3$& $-77.9$&$-64.4$\\
& 11819.560& $-68.3$& $-79.2$&$-66.7$\\
& 11821.561& $-72.0$& $-80.7$&$-69.3$\\
\hline
\end{tabular}
\end{minipage}
\end{table*}
\begin{table}
\caption {Adopted rest wavelengths of the lines used to compute the RVs displayed in Figs.\,\ref{fig perhut}, \ref{fig perasl} and \ref{fig RV} and listed in Table\,\ref{tab rv}.\label{tab wavel}} 
\begin{tabular} {l r c}
\hline
Ion & Effective &  Nature\\
& wavelength & \\
\hline\hline
\hee\ & 4199.830 & Abs. \\
\hee\ & 4541.590 & Abs. \\
\ss\  & 4552.654 & Em. \\
\ss\  & 4567.872 & Em. \\
\nn\  & 4634.250 & Em. \\
\nn\  & 4641.020 & Em. \\
\hee\ & 4685.682 & Em. \\
\hline
\end{tabular}
\end {table}

\subsection{Short-term variability}
\begin{figure}
\resizebox{\hsize}{!}{\includegraphics{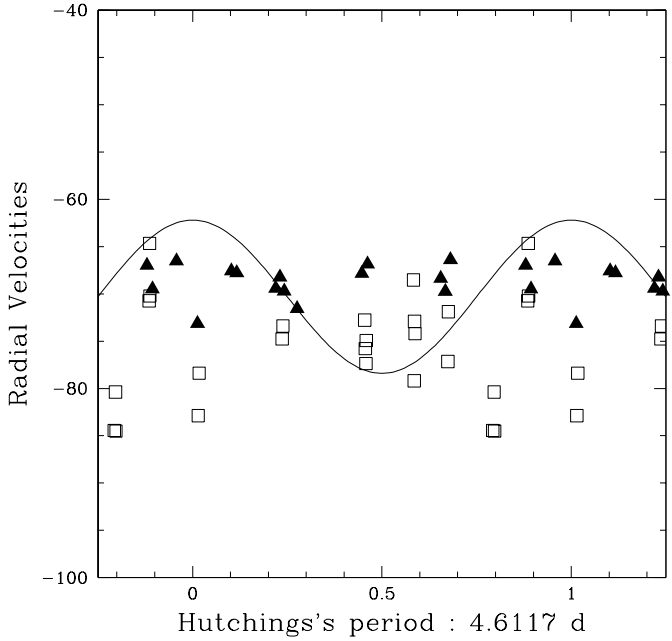}}
\caption{ \label{fig perhut} RVs of \hee $\lambda$\,4200 as measured on our 1987 spectra (open squares) and RVs of \hee $\lambda$\,4542 from our 1998 data (filled triangles) folded with Hutchings' period of 4.6117 d and superimposed on his orbital solution.}
\end{figure}

\begin{figure}
\resizebox{\hsize}{!}{\includegraphics{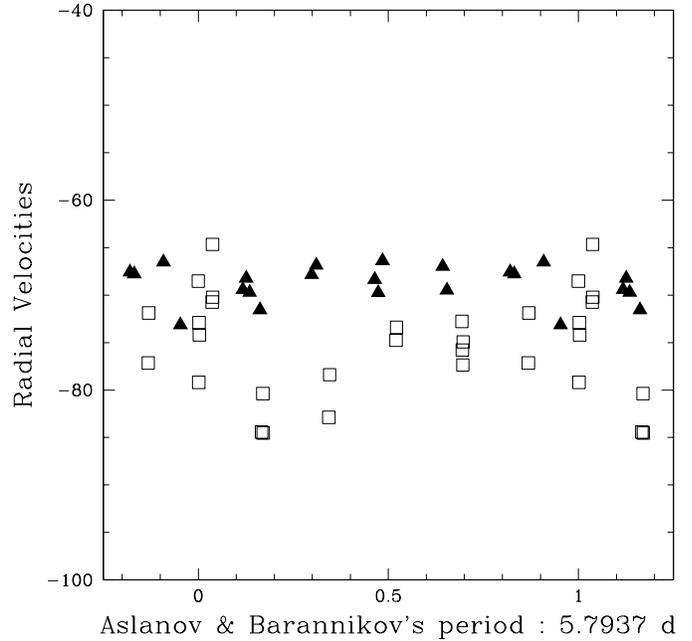}}
\caption{ \label{fig perasl} Same as Fig. \ref{fig perhut}, but for the period of 5.7937 d proposed by Aslanov \& Barannikov (\cite{asl}).}
\end{figure}

Over all those years of intense study of \hd, only a few authors really addressed the issue of short-term variability of this star, even if most of the periods found were rather short. For instance, Vreux \& Conti (\cite{vr}) obtained two spectra per night during two consecutive days of their observing run. However, to analyse completely the short-term variability, more spectra taken on numerous consecutive days are needed. To investigate this problem, we collected at least two spectra per night during the 7 days of the August 1987 run and during the 6 days of the November 1998 run. The 1987 data cover the \hee $\lambda$\,4200 line, while the 1998 data cover the \hee $\lambda$\,4542 line. These data enable us to search for short-term variations of the spectra, and also to check the short periods (a few days) given in the literature. Because of the poorer spectral resolution (1.7\,\AA) of the 1987 data, the RVs show a large scatter in 1987 compared to the 1998 measurements that were obtained with a resolution of 0.2\AA\ (see Figs.\,\ref{fig perhut} and \ref{fig perasl}).\\
We tried to search for a periodicity in the \hee RVs measured on the two sets of data (20 spectra in 7 days + 15 spectra in 6 days), but we did not find any significant short period. Moreover, those data do not confirm the periods given in the literature. In Figures \ref{fig perhut} and \ref{fig perasl}, our data are folded respectively with Hutchings'period (4.6117 d, Hutchings \cite{hut1}) and the 5.7937\,days period of Aslanov \& Barannikov (\cite{asl}). Both periods can clearly be ruled out. Due to the severe aliasing of our time series, the 1.02\,day period taken from Vreux \& Conti (\cite{vr}) cannot be sampled over an entire cycle with our data. Though the scatter of the folded data (20 km s$^{-1}$ for $\Delta\phi\sim$0.1 in 1987 and 10 km s$^{-1}$ for $\Delta\phi\sim$0.3 in 1998) argues against this last period, it cannot be totally excluded. Meanwhile, we emphasize that no short period, typically from 1 to 5 days, does appear in our data. \\
We have also applied the period research techniques to the \hee EWs determined in those two runs. Again, no significant period appears. With the 1998 high-resolution data, we also searched for short-term variations in the line profiles: neither \he $\lambda$\,4471 nor \hee $\lambda$\,4542 seem to change during the whole duration of our run. Short term variations of typically 1 to 5 days - if they exist - are thus only of very small amplitude.

\subsection{Long-term behaviour}
\begin{figure}
\resizebox{\hsize}{!}{\includegraphics{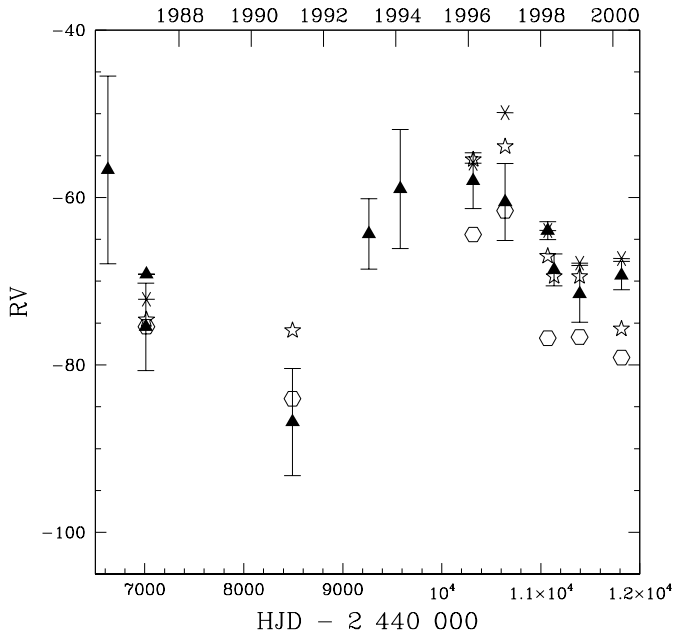}}
\caption{ \label{fig RV} Mean RVs measured for each observing run: there are \hee $\lambda\lambda$\,4200, 4542 absorptions (represented by filled triangles) and some emission lines like \ss $\lambda\lambda$\,4552, 4568 (stars), \nn $\lambda\lambda$\,4634, 4641 (hexagons) and \hee $\lambda$\,4686 (asterisks). The bottom axis yields the date in heliocentric julian days (HJD), while the x-axis on top of the panel indicates the year of observation: tickmarks represent the middle of the year quoted above. 1-$\sigma$ error bars are shown for \hee $\lambda\lambda$\,4200, 4542 RVs (when no error estimates are shown, it means that only one spectrum was available). }
\end{figure}

\begin{figure}
\resizebox{\hsize}{!}{\includegraphics{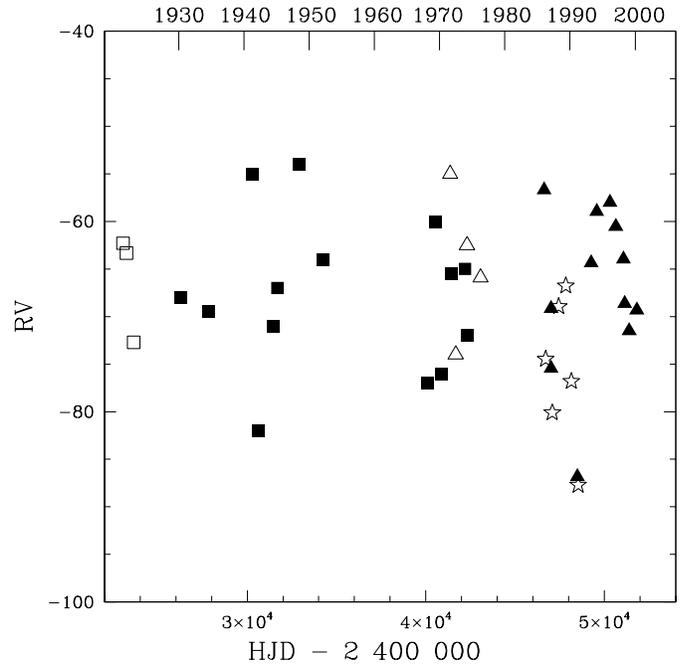}}
\caption{ \label{fig RVtot} All published radial velocities of absorption lines (mainly \hee $\lambda$\,4200, 4542): the open squares represent the mean data from Plaskett (\cite{plask}), the filled squares from Hutchings (\cite{hut1}), the open triangles from Vreux \& Conti (\cite{vr}), the stars from Underhill (\cite{und2}) and the filled triangles from this work.}
\end{figure}

\begin{figure}
\resizebox{\hsize}{!}{\includegraphics{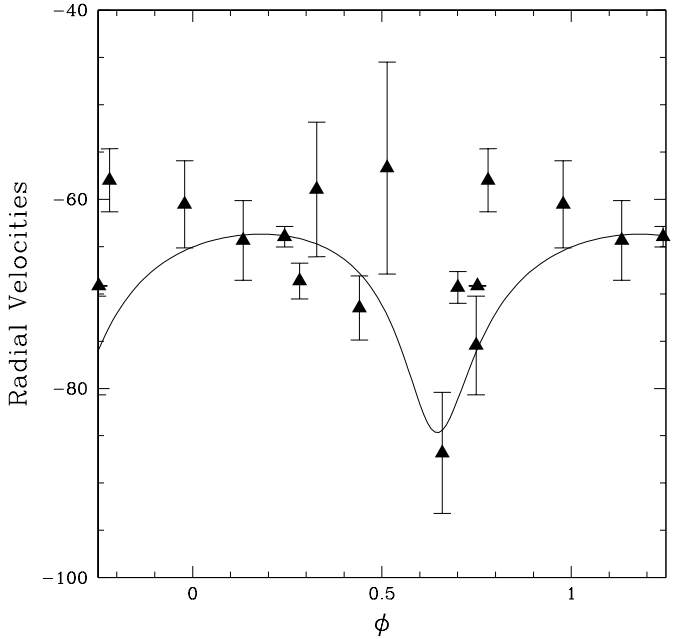}}
\caption{ \label{fig perbar} Our mean \hee data folded with Barannikov's (\cite{bar}) period of 1627.6 d, superimposed on his orbital solution. 1-$\sigma$ error bars are shown:  when no error estimates are shown, it means that only one spectrum was available.}
\end{figure}

Since we do not detect any short-term variability in the \hee RVs, we can average the \hee absorption RVs for each observing run: we have plotted those means in Fig. \ref{fig RV}, and we have combined these values with all published RVs in Fig. \ref{fig RVtot}. We caution that the oldest data are rather uncertain, and that they should only be used for a qualitative analysis.\\  
The \hee $\lambda\lambda$\,4200, 4542 lines are most probably formed in the deeper (photospheric) layers of the stellar atmosphere and can thus give rather unpolluted information about the star's actual motion. Other absorption lines (Mg~{\sc{ii}}\ $\lambda$\,4481, \si $\lambda$\,4631 and all \nn lines between 4510 and 4534 \AA) and some emission lines (the unidentified Of emissions $\lambda\lambda$\,4486, 4504 ; \ss $\lambda\lambda$\,4552, 4568 ; \nn $\lambda\lambda$\,4634, 4341 ; \hee $\lambda$\,4686 and \od $\lambda$\,4705) display the same behaviour as the \hee absorption lines. The RVs of some of these lines are shown in Fig. \ref{fig RV}. On the other hand, the RVs of the main absorption components of H$\beta$, H$\gamma$ and \he $\lambda\lambda$\,4388, 4471, 4713 become less and less negative over all the years of our observing campaign. This is linked to the long-term line profile variations discussed in Section 6.\\
The continuous trend of the absorption lines towards more negative RVs since 1922-1923, as reported by Underhill\footnote{The errors quoted in Table 3 of Underhill (\cite{und2}) are not the standard deviations of the data, contrary to what we used in Fig \ref{fig RV} or \ref{fig perbar}, but the standard deviations of the means: those two values differ by a factor $\sqrt{N}$. This explains the rather low value of the $\sigma$ in her Table.} (\cite{und2}), does not appear in our data. On the contrary, in 1994 and in 1996, the RV of \hee $\lambda\lambda$\,4200, 4542 is about the same as in 1922. The absence of a clear very long term trend is best seen in Fig.\,\ref{fig RVtot}.\\
Underhill (\cite{und2}) also mentionned that the RVs of the absorption and emission lines show different behaviours. But in our data, \hee absorptions and He~{\sc ii}, O~{\sc ii}, \si and \nn emissions display quite similar behaviours (as can be seen on Fig. \ref{fig RV} and on Fig. 1 of Barannikov \cite{bar}). However, the `abrupt' shift of the RVs towards more negative values in 1991, seen by Underhill (\cite{und2}) does appear in our data as well as in those of Barannikov (\cite{bar}). If intrinsic to a single star, this behaviour could indicate an episodic change of the velocity structure in the inner regions of the expanding atmosphere.\\
Using our data set only, a period of about 4600 days appears in the Fourier periodogram. However, with $SL = 0.89$, this detection is not significant. Moreover, this period completely disappears when we include the RVs determined by Plaskett (\cite{plask}), Hutchings (\cite{hut1}), Vreux \& Conti (\cite{vr}), Underhill (\cite{und2}) and Barannikov (\cite{bar}). Considering the whole data set, no significant period becomes visible, the minimum $SL$ reaching $0.93$.\\
According to Barannikov (\cite{bar}), the \hee RVs should follow a period of 1627.6 d. But when we fold our data according to this period, it clearly appears that it is ruled out by our data (see Fig. \ref{fig perbar}). Barannikov used low-resolution\footnote{The dispersion was 44 \AA/mm$^{-1}$.} photographic spectra, which are prone to errors (there are differences up to 70 km\,s$^{-1}$ between two consecutive nights for a 1627.6 days period with an amplitude K of only 10.5 km\,s$^{-1}$!): this might explain the important scatter seen in his orbital solution. We note that the minimum in Barannikov's radial velocity curve occurred in 1991, i.e.\ at the same epoch as we observe the RV shift discussed hereabove. However, our data do not suggest that this phenomenon is periodic.

\section{Equivalent Widths}
\begin{figure}
\resizebox{\hsize}{!}{\includegraphics{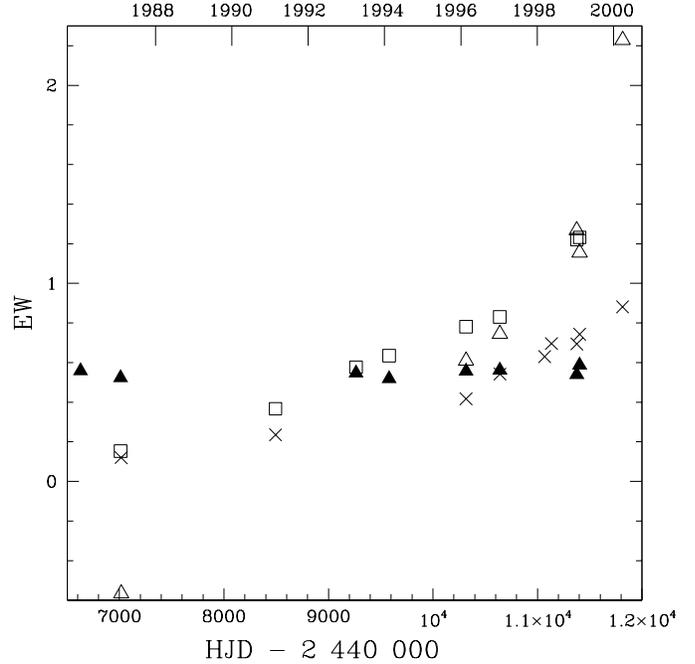}}
\caption{ \label{fig ew} Average EW for each observing run: \hee $\lambda$\,4200 is represented by filled triangles, H$\gamma$ by open squares, \he $\lambda$\,4471 by crosses and H$\beta$ by open triangles. The errors are very small, roughly the size of the symbols used in the figure, and they are listed in Table \ref{tab ewtab}.}
\end{figure}

Aslanov \& Barannikov (\cite{asl}) claim that the equivalent width variability of emission lines is of stochastic nature, while the one associated with absorption lines exhibits a period twice the one derived from their radial velocities. More recently, Underhill (\cite{und2}) found no periodicity in the variability of the EWs.\\
In our data, two different categories of lines emerge: most of the lines appear rather constant but several others have clearly variable EWs. In the first category, \si $\lambda$\,4088, \hee $\lambda$\,4200, 4542 and \he $\lambda$\,4143 exhibit maximum EW variations below $\sim 20$\% and the scatter is quite low (see Table \ref{tab ewtot}). Other lines fall also into this category, e.g.\ the \od and \cd emission lines, the unidentified Of emissions and the \nn absorptions. In agreement with Underhill (\cite{und2}), we do not find any significant periodicity when considering the variability of the EWs of the \hee absorption lines. \\
Turning now to the lines showing variable EWs, we will focus here on the behaviour of those lines for which we have the best coverage with our data sets. The \ss $\lambda\lambda$\,4552, 4568 and \cc $\lambda\lambda$\,4647-50 emissions seem to decrease gradually during the whole duration of our observing campaign. The EWs are reduced by a factor 2 between 1987 and 2000. The \he $\lambda$\,4713 line also presents EW variations, up to 50\% from the mean. But the most impressive changes occur in \hg, \he $\lambda$\,4471 and \hb. Their total equivalent widths (absorption + emission) increase gradually (see Table \ref{tab  ewtab}). In Fig. \ref{fig ew}, one can clearly see this continuous trend to higher EWs, as the emission component weakens and finally disappears (see Section 6). \\
This separation of the lines in 2 categories, the rather constant ones (e.g. \hee emission and absorptions) and the clearly variable ones (e.g. \ss and the most variable \he and H~{\sc{i}} lines), points most probably towards different formation regions for each category.

\begin{table}[htb]
\caption { \label{tab ewtot} Mean EW for different lines, averaged over the entire observing campaign. The quoted EWs for \hb\ and \hg\ are the total ones, including the \hee Pickering absorption (and the DIB in the case of \hb). The last two columns summarize the results of a TVS analysis (see Section 6). The $\sqrt{TVS}$ is given in units of the continuum flux, while the last column provides the wavelength interval over which the detected variations are significant at the 99\% level.} 
\begin{tabular} {l c c c }
\hline
\vspace*{-3mm}\\
Line& EW(\AA) &$\sqrt{TVS}$& width(TVS)\\
& (\AA)& &(\AA)\\
\hline\hline
\si$\lambda$4088&0.196$\pm$0.021&&\\
\he$\lambda$4143&0.040$\pm$0.006&&\\
\hee$\lambda$4200&0.538$\pm$0.033&&\\
\hg&0.506$\pm$0.385&&\\
\he$\lambda$4471&0.643$\pm$0.215&0.065&4462.3-4473.3\\
\hee$\lambda$4542&0.672$\pm$0.057&0.017&4535.5-4542.6\\
\ss$\lambda$4552&-0.141$\pm$ 0.045&0.016&4549.9-4555.3\\
\ss$\lambda$4568&-0.110$\pm$ 0.030&0.010&4565.4-4569.4\\
\nn$\lambda$4634&-1.040$\pm$0.059&0.014& 4630.7-4637.0\\
\nn$\lambda$4641& &0.018&4637.0-4644.4\\
\cc$\lambda\lambda$4647-50&-0.306$\pm$0.077&0.013&4644.4-4652.7\\
\hee$\lambda$4686& -0.524$\pm$0.045&0.013&4709.2-4714.2\\
\he$\lambda$4713&0.134$\pm$0.040&0.092&4849.0-4866.5\\
\hb&1.457$\pm$0.788&\\
\hline
\end{tabular}
\end {table}

\begin{table*}[htb]
\caption { \label{tab ewtab} Average EW for each observing run for \hee $\lambda$\,4200, H$\gamma$, \he $\lambda$\,4471 and H$\beta$. Standard deviations are also quoted : when no error estimates are given, it means that only one spectrum was available.} 
\begin{center}
\begin{tabular} {l l l l l}
\hline
\multicolumn{1}{c}{Run}& \multicolumn{1}{c}{EW \hee} & \multicolumn{1}{c}{EW H$\gamma$} & \multicolumn{1}{c}{EW \he} & \multicolumn{1}{c}{EW H$\beta$}\\
\hline\hline
Jul. 1986& $0.559 \pm 0.003$ & &  & \\
Jul. 1987& $0.524 \pm 0.019$ &$0.153 \pm 0.038$ & & \\
Aug. 1987 & & & $0.120$&$-0.566$ \\
Aug. 1991&  &$0.367 \pm 0.034$ &$0.235 \pm 0.019$ & \\
Oct. 1993&$0.548 \pm 0.051$ &$0.577 \pm 0.075$ & & \\
Aug. 1994&$0.518 \pm 0.012$ &$0.635 \pm 0.017$ & & \\
Aug. 1996&$0.557 \pm 0.049$ &$0.781 \pm 0.017$ &$0.417 \pm 0.017$ &$0.610 \pm 0.066$\\
Jul. 1997&$0.562 \pm 0.033$ &$0.830 \pm 0.045$ &$0.542 \pm 0.006$ &$0.745 \pm 0.018$\\
Sep. 1998&  & &$0.622 \pm 0.029$ & \\
Nov. 1998&  & &$0.696 \pm 0.039$ & \\
Jul. 1999&$0.539 \pm 0.018$ &$1.221 \pm 0.055$ &$0.694 \pm 0.021$ &$1.267 \pm 0.070$ \\
Aug. 1999&$0.587 \pm 0.057$ &$1.232 \pm 0.043$ &$0.743 \pm 0.017$ &$1.155 \pm 0.045$ \\
Sep. 2000& & & $0.881 \pm 0.036$ & $2.229 \pm 0.090$\\
\hline
Mean& $0.538 \pm 0.033$& $0.506 \pm 0.385$& $0.643 \pm 0.215$& $1.457 \pm 0.788$\\
\hline
\end{tabular}
\end{center}
\end {table*}

\section{Line Profile Variability}
As summarized for example in Andrillat et al.\ (\cite{and}), many line profiles of the spectrum of \hd\ vary over the years, with some of them changing spectacularly from P Cygni profiles to pure absorption or emission lines. This is especially the case for the Balmer Hydrogen lines and some \he lines ($\lambda\lambda$\,4388, 4471, 4713). In our data, \he $\lambda\lambda$\,4388, 4713 appeared only once, in 1987, with a P Cygni profile while these lines display pure absorptions since then.\\
To better quantify the line variability, we used the temporal variance spectrum (TVS), as defined by Fullerton et al.\ (\cite{ful}). This TVS analysis, when applied on the common spectral range of the data collected in 1987 and from 1996 to 2000, yields interesting results (see Table \ref{tab ewtot}): in addition to the \he and  H~{\sc{i}} lines, the absorption lines of \nn (from 4510\AA\ to 4534\AA) and of \hee $\lambda$\,4542, as well as the emission lines of \nn $\lambda\lambda$\,4634-41, \hee $\lambda$\,4686, \ss $\lambda\lambda$\,4552, 4568 and \cc $\lambda\lambda$\,4647-50, do all vary, though with a smaller amplitude. On the contrary, other emission lines (e.g. \od and the unidentified $\lambda\lambda$\,4486, 4504) were stable during the whole observing campaign.\\
To compare the variations of the different lines, we have used the local pattern cross-correlation technique discussed by Vreux et al.\ (\cite{vr2}) and by Gosset et al.\ (\cite{gos}). But before that, we have convolved the 1998 and 2000 data until they reach the same resolution as the 1996, 1997 and 1999 spectra. Only then, we have cross-correlated the variation pattern of some unblended lines (e.g.\ \he $\lambda$\,4471, \hee $\lambda$\,4686) with the whole spectra. Using the \hee $\lambda$\,4686 emission line as a reference, a strong correlation appears between the deformation pattern of \hee $\lambda$\,4686 and the ones of the \nn $\lambda\lambda$\,4634, 4641 lines. On the contrary, the \he $\lambda$\,4471 pattern seems rather related to the variation of the \nn absorptions between 4510 and 4534\AA, \hee $\lambda$\,4542, \cc $\lambda$\,4647-50, \he $\lambda$\,4713 and \hb. The behaviour of the \nn $\lambda\lambda$\,4634-41 lines suggests therefore that these lines form in the same physical region as the \hee $\lambda$\,4686 emission, while the clearly distinct variations of the \he and Balmer emissions point towards a different origin and/or emission mechanism for the latter lines. The variability pattern of \hee $\lambda$\,4686 and \nn $\lambda\lambda$\,4634-41 between 1996 and 2000 reveals an enhancement of the core of the line while the red wing is progressively depleted.\\
These \nn profile variations probably point towards a wind contribution to the emission. In this context, it is worth recalling that once that a velocity gradient exists, the fluorescence mechanism, suggested by Swings (\cite{Swings}) to account for the occurrence of the \nn $\lambda\lambda$\,4634-41 emission lines in the spectra of Of stars, can become extremely effective (Mihalas \cite{Mihalas}).\\ 
\begin{figure}
\resizebox{\hsize}{!}{\includegraphics{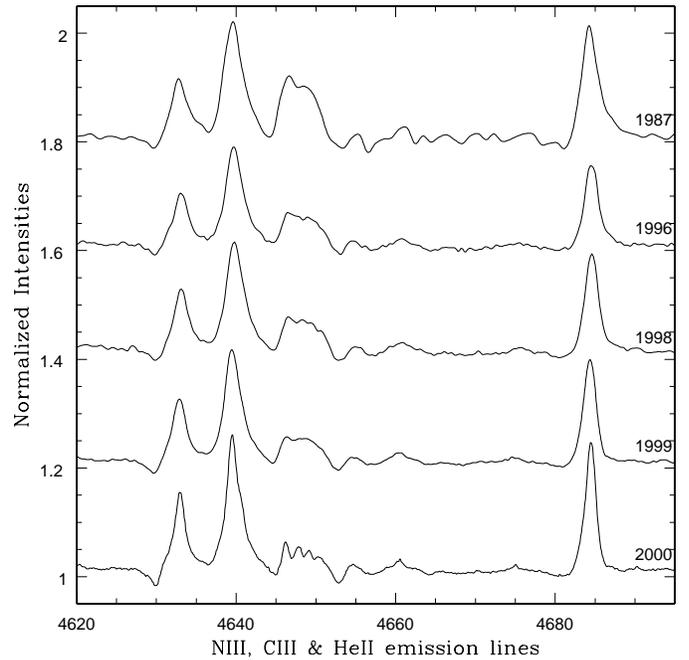}}
\caption{ \label{fig raies2} Aspect of \nn $\lambda\lambda$\,4634-41, \cc $\lambda\lambda$\,4647-50 and \hee $\lambda$\,4686 for different years: the dispersion used in 1987, 1996 and 1999 is 33 \AA\ mm$^{-1}$, while it is 16  \AA\ mm$^{-1}$ in 1998 and 2000 (see Table \ref{tab obs}). The \cc lines show a very significant decrease in intensity between 1987 and the recent years, while all lines seem sharper in 2000. The heliocentric correction has not been applied to the spectra displayed on this figure.}
\end{figure}
The most impressive variability is unquestionably displayed by \ha, \hb, \hg\ and \he $\lambda$\,4471, as already quoted in Section 5. The long-term line profile variability of these lines is presented in Table\,\ref{tab var} and Fig.\,\ref{fig raies}: the changes are clearly outstanding. We do not have a lot of information about \ha, but the line is clearly variable. Our 1997 \ha\ data yield a normalized peak intensity of 1.8, whereas in 1990 (Underhill \cite{und2}), it reached a value of 2.5. A similar behaviour was observed by Beals (\cite{bea}): in 1938, the normalized intensity of \ha\ was over 2, but it was only 1.3 in 1945. The \ha\ line thus roughly behaves in the same way as \hb, \hg\ and \he $\lambda$\,4471. Other Balmer lines (e.g. H$\delta$, H$\epsilon$) and \he lines (e.g.\ 4388, 4713) also display a similar behaviour, but we will now focus on the \hg, \he $\lambda$\,4471 and \hb\ lines, for which we have the most extensive data set.\\
These three lines displayed a P Cygni profile at the beginning of our observations in 1987. Then the emission component progressively weakened until the lines appeared completely in absorption. The first line that went into absorption is \he $\lambda$\,4471, followed by H$\gamma$ and finally H$\beta$. After this change, the depth of the \he absorption lines continued to increase gradually, as can be seen in Figs.\,\ref{fig ew} and \ref{fig raies}. Since the change into absorption occurred only recently, we have no information yet about the behaviour of the \hg\ and \hb\ lines after the emission has disappeared.\\ Interestingly, a similar transition was observed some 56 years ago (see Table \ref{tab var}): all three lines were seen as pure absorptions in 1944 and also in 2000. In 1934, \he $\lambda$\,4471 was already described as an absorption, while H$\gamma$ and H$\beta$ were still displaying P Cygni profiles. Then, \hg, followed very soon by \hb, went into absorption. Finally, in 1954, a weak emission reappeared in \hb. Therefore, the long term line profile variations of \hd\ seem to be recurrent with a timescale of about 56 years.\\
However, the interpretation of the Balmer line profile variations is complicated by the fact that these lines are blended with \hee Pickering lines. Following the same procedure as in Lamers \& Leitherer (\cite{lam}), we attempted to restore the H~{\sc{i}} line profiles by subtracting a fake \hee line obtained by interpolation (\hg) and extrapolation (\hb) from the unblended \hee $\lambda$\,4200 and $\lambda$\,4542 lines. The reconstructed profiles are striking: the hydrogen lines no longer appear as P Cygni profiles, but as rather sharp and symmetrical emissions superimposed on broader absorptions (see the lower panels of Fig. \ref{fig raies}). This emission fades with time, and finally nearly disappears in \hg\ in 1999. The RVs of the `restored' \hb\ and \hg\ emission peaks become more negative during the last two years.\\
We have also tried to remove the photospheric \he absorption to recover the stellar wind line profile using two slightly different approaches. First, we assume that the emission has completely disappeared in the spectra observed in 2000, and use this profile for the \he photospheric contribution. Second, we evaluate the photospheric absorption by fitting a gaussian profile to the blue wing of the $\lambda$\,4471 line, assuming it is not contaminated by emission. In this latter case, we set the RV of the photospheric \he component to be equal to the observed RV of the \hee $\lambda\lambda$\,4200, 4542 lines. The resulting pure emission profiles are displayed in Fig.\,\ref{fig raies}. We caution however that the reconstruction of the \he $\lambda$\,4471 wind profile is far more uncertain than the restoration of Balmer lines.\\
The rather narrow Balmer emission profiles without evidence for an associated P Cygni wind absorption are probably difficult to explain in a spherically symmetric model for the wind of \hd. In fact, the width of the lines indicates that they must be formed over a small range in radial velocity and hence the radial extent of the line forming region in the expanding wind should be rather small. Under such circumstances, a spherically symmetric wind would most probably result in a pronounced P Cygni absorption component that is not observed.\\
On the other hand, the different behaviour of the \hee $\lambda$\,4686 and \nn emission lines on the one side and the \he and Balmer lines on the other side, clearly indicates that there must be at least two distinct regions in the atmosphere of \hd\ where emission lines are formed.

\begin{figure*}
\resizebox{\hsize}{!}{\includegraphics{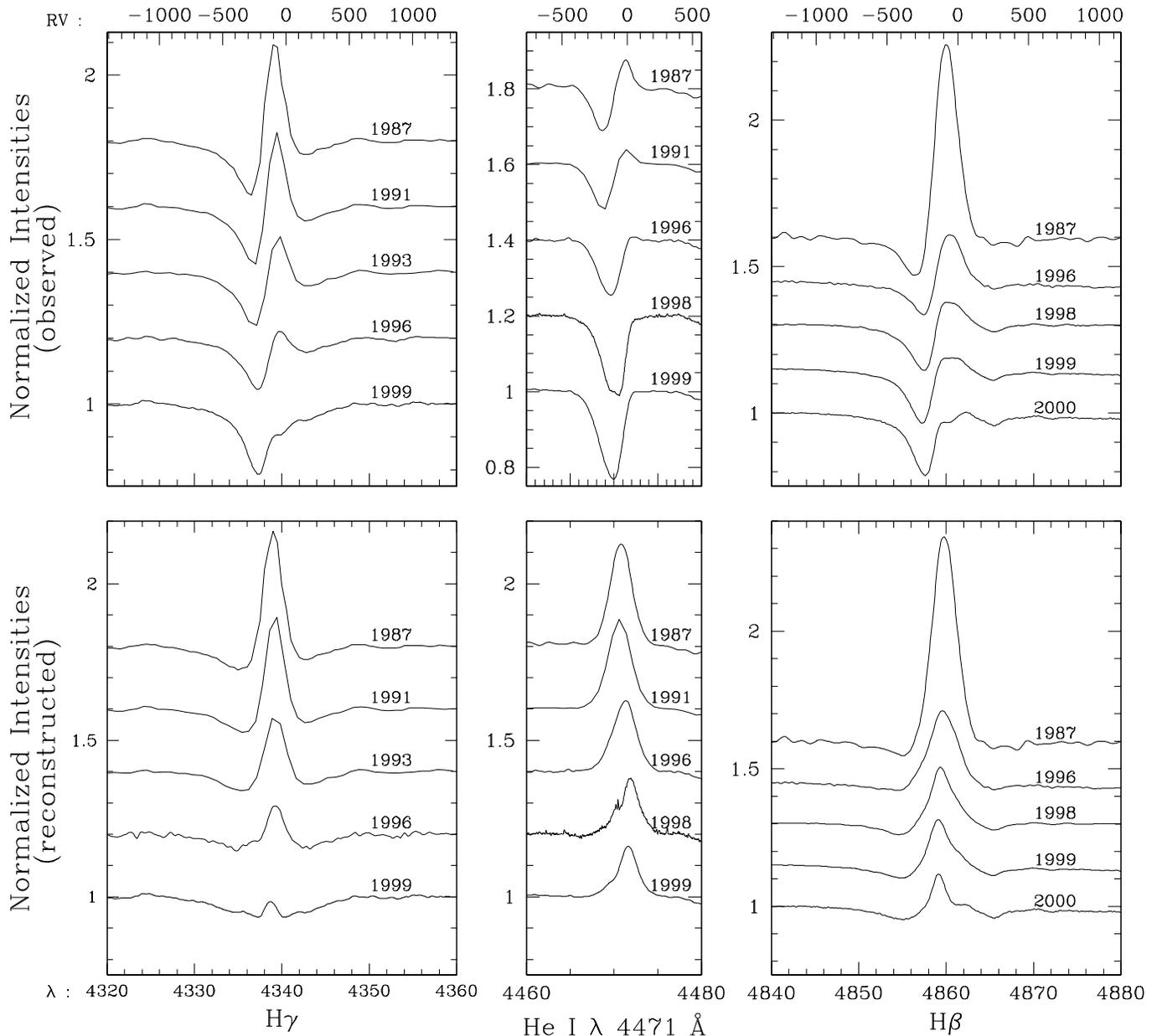}}
\caption{ \label{fig raies} Upper panels: aspect of H$\gamma$, \he $\lambda$\,4471 and H$\beta$ for some recent years. The lines gradually evolve from P Cygni profiles into pure absorption lines. Bottom panels: reconstructed profiles of the same lines (see text). The \hee contribution has been subtracted from \hg\ and \hb\ (bottom, left and right) while the \he photospheric component has been subtracted from the \he total profile (bottom, middle). The heliocentric correction has not been applied to the spectra displayed in this figure.}
\end{figure*}

\begin{table}
\caption { \label{tab var} Aspect of the H$\beta$, H$\gamma$ and \he $\lambda$\,4471 lines since 1919. The codes for the references are the following:  Pl24 for Plaskett (\cite{plask}), Me25 for Merrill et al.\ (\cite{mer}), Sw42 for Swings \& Struve (\cite{swi}), Be50 for Beals (\cite{bea}), Ma55 for Maninno \& Humblet (\cite{man}), Ho68 for Houziaux \& Ringuelet-Kaswalder (\cite{hou}), An73 for Andrillat et al.\ (\cite{and}), Hu75 for Hutchings (\cite{hut1}), Un94 for Underhill (\cite{und2}) and Ba99 for Barannikov (\cite{bar}). } 
\begin{tabular} {l r r r r}
\hline
 Date & H$\beta$ & H$\gamma$ & \he & reference\\
\hline\hline
1919& P Cyg.&P Cyg. & P Cyg.& Me25\\
1921-1923& P Cyg.& P Cyg.& &Pl24\\
1934-1938&P Cyg. & P Cyg.&Abs. &Be50\\
1941& P Cyg.& P Cyg.& Abs.&Ma55\\
1942&P Cyg. & P Cyg.& &Sw42\\
& & & &Ma55\\
1944-1945& Abs.& Abs.&Abs. &Be50\\
1950& Abs.& Abs.&Abs. &Ma55\\
1953& Abs.& Abs.& Abs.&Ma55\\
1954& P Cyg.& Abs.& Abs.&Ma55\\
&(weak)& & & \\
1966& P Cyg.& & &Ho68\\
1968-1974& P Cyg.&P Cyg. & P Cyg.&An73\\
& & & &Hu75\\
1976& P Cyg. & P Cyg.& P Cyg.&this work\\
1982-1985&P Cyg. & P Cyg.& &Ba99\\
1986-1991&P Cyg. &P Cyg. & P Cyg.&Un94\\
1989-1991& P Cyg.& P Cyg.& &Ba99\\
1991& & P Cyg.& P Cyg.& this work\\
1993-1994& & P Cyg.& &this work\\
1996-1997& P Cyg.& P Cyg.& Abs.&this work\\
1998-1999& P Cyg.& Abs.& Abs.&this work\\
2000&Abs.& &Abs.&this work\\
\hline
\end{tabular}
\end {table}

\section{Discussion}
The spectrum of \hd\ looks pretty much like that of the extreme O8\,Iaf star HD\,151804 (Crowther \& Bohannan \cite{crow}), but exhibits additional emission lines (C~{\sc ii}, O~{\sc ii},...) and presents spectacular H~{\sc{i}} and \he line variations.\\
Underhill (\cite{und2}) found similarities between the spectrum of \hd\ and the spectra of Ofpe/WN9 stars (Bohannan \& Walborn \cite{boh}) and those of B[e] stars. Comparing with more recent data (e.g.\ Nota et al.\ \cite{nota}), we notice some marked differences between the spectrum of \hd\ and spectra of Ofpe/WNL stars. For example, the \hee and \nn absorptions are stronger in the spectrum of \hd. Also, the \ha\ emission of \hd\ is sharper and the \he lines can evolve into pure absorptions, not only weaker P Cygni.\\ 
\hd\ was also discussed in connection to the Luminous Blue Variable (LBV) stars. However, until now, \hd\ has not presented spectacular spectral type variations. The \hee $\lambda$\,4542 line is only slightly variable and the apparent changes in spectral type discussed in Sect.\,3.1 rather reflect the variations of the wind emission in \he $\lambda$\,4471. Finally, unlike most LBVs or LBV candidates, no nebular emission was detected around \hd, even if the star has been suggested in the past to be the exciting star of a faint nebula (see Higgs \& Ramana \cite{hig}). But, in a more recent work Lozinskaia (\cite{loc}) excludes the presence of a nebula around \hd.\\

Andrillat et al.\ (\cite{and2}) presented some characteristics of Oe stars. In their sample of stars, \ha, \he $\lambda$\,6678 and H~{\sc{i}} Paschen lines are often present in emission. These emissions are variable, on time scales of several years. They can completely disappear or even become absorption lines. \hd\ presents similar characteristics of varying \he and H~{\sc{i}} lines, and its near IR spectrum is also similar to those of Oe stars. \hd\ does not display the Fe~{\sc{ii}} $\lambda\lambda$\,7515, 7712 and O~{\sc{i}} $\lambda$\,7772 lines seen by Andrillat et al.\ (\cite{and2}) in most Oe spectra, but these lines are also missing in the spectra of some confirmed Oe/Be stars, as for instance in X Persei.\\
Moreover, Divan et al.\ (\cite{div}) showed that Oe and Be stars are brighter and redder when the emissions in the Balmer continuum and in \ha\ are stronger. Unfortunately, there are few photometric studies of \hd. Its V magnitude was measured by Plaskett \& Pearce (\cite{pla}), Hiltner \& Johnson (\cite{hilt}), Blanco et al.\ (\cite{bla}) and Leitherer \& Wolf (\cite{lei}) who all found V=7.40 and by Bouigue et al.\ (\cite{bou}), who found V=7.35. The B-V color index is about 0.18 for all articles, except the last one (B-V = 0.14). However, \hd\ was sometimes attributed a variable character: in the New Catalogue of Possible Variable Stars (NVS), its V magnitude varies from 7.35 to 7.48. Chilardi et al.\ (\cite{chi}) reported magnitude variations up to 0.4 mag, and more recently, Barannikov (\cite{bar}) showed that there was an abrupt change in the light curve in 1996 and 1997, when \he $\lambda$\,4471 went into absorption, followed quite soon by H$\gamma$. At that time, the star became suddenly fainter and bluer (in $\Delta$(V-R)). In addition, the Hipparcos satellite has measured a mean H$_p$ magnitude of 7.412$\pm$0.012 mag with a maximum of 7.39 and a minimum of 7.43 (Hipparcos main catalogue, ESA 1997). The small amplitude of the light curve argues against \hd\ being a dormant LBV, but it could be compatible with an Oe/Be behaviour.\\

\hd\ seems thus to share several characteristics of Oe or Be-type stars. However, the spectrum of \hd\ displays rather sharp emissions without any indication of a double-peaked morphology that could be attributed to an equatorial disc or flattened wind. Therefore, if \hd\ has indeed an equatorial disc, it should be seen under a rather low inclination. In fact, spectropolarimetric observations (Harries \cite{Harries}) do not exclude the possibility of a disc, provided it would be seen face-on.\\
Moreover, as mentioned in the introduction, several conflicting determinations of the mass-loss rate of \hd\ can be found in the literature. From the unsaturated resonance lines seen in the IUE spectrum of \hd, Hutchings \& von Rudloff (\cite{HvR}) and Howarth \& Prinja (\cite{HP}) inferred rather low values of $2 \times 10^{-7}$ and $3 \times 10^{-6}$\,M$_{\odot}$\,yr$^{-1}$ respectively. Using the H$\alpha$ line flux, Peppel (\cite{Pep}) and Leitherer (\cite{Leit}) obtained $\dot{\rm M} \sim 10^{-5}$\,M$_{\odot}$\,yr$^{-1}$. Finally, the largest mass-loss rates ($5 \times 10^{-5}$\,M$_{\odot}$\,yr$^{-1}$ comparable to those of Wolf-Rayet stars) were derived from the infrared excess (Ferrari-Toniolo et al.\ \cite{fer}). The various values listed here refer to observations obtained at the same epoch (between 1978 and 1982) thus ruling out any long-term trend as the origin of the discrepancies. These different determinations of $\dot{\rm M}$ could provide further support for a scenario based on a disc seen face-on. In fact, the UV resonance lines are sensitive to the absorption along the line of sight, whereas the IR continuum probes the emission from the extended disc.\\

As already quoted in the first section, \hd\ has been sometimes classified as a runaway: Bekenstein \& Bowers (\cite{bekbow}) found a peculiar velocity $v_p$ $>$ 98 km\,s$^{-1}$ while Underhill (\cite{und2}) gives $v_p = 43$\,km\,s$^{-1}$. The Hipparcos satellite has measured the proper motion of \hd\ and found $\mu_{\alpha\,cos(\delta)}= (-5.39\pm 0.66) \times 10^{-3}$\,arcsec\,yr$^{-1}$  and $\mu_{\delta}= (-0.93\pm 0.54) \times 10^{-3}$\,arcsec\,yr$^{-1}$. Using the method described by Moffat et al.\ (\cite{mof} and erratum in \cite{moferr}), we calculate : \\
$(\mu_l)_{pec}= (-3.14\pm1.01) \times 10^{-3}$\,arcsec\,yr$^{-1}$\\
$(\mu_b)_{pec}= (0.47\pm0.56) \times 10^{-3}$\,arcsec\,yr$^{-1}$\\
yielding $(v_t)_{pec} = 37.8 \pm 16.5$\, km\,s$^{-1}$, assuming a distance from the Sun equal to 2.51 kpc (Gies \cite{gies}) and a 30\% uncertainty in the distance.\\
Moffat et al.\ (\cite{mof}) define a runaway as a star with $(v_t)_{pec} > 42 + \sigma_{(v_t)_{pec}}$. This criterion does not allow us to attribute a definite runaway status to \hd.

\section{Conclusions}
Using a 30 years long campaign of spectroscopic observations, we have analysed the strange behaviour of \hd. The analysis of our RV time series failed to reveal a significant period and our data rule out the periods claimed in previous studies (e.g.\ Hutchings \cite{hut1}, Barannikov \cite{bar}). However, \hd\ presents some outstanding long-term variations: \he and H~{\sc{i}} lines changed recently from P Cygni profiles to apparently pure absorption lines. A similar situation was observed about 56 years ago. The line profile variations are thus very probably recurrent. We tried to remove the underlying photospheric absorption to better understand some of the variations. The restored Balmer and \he lines consist of rather narrow emission lines overlying a broad (photospheric) absorption line and reveal no evidence for a P Cygni type absorption. The other emission lines either display lower amplitude variations (\nn\ $\lambda\lambda$\,4634-41, \hee\ $\lambda$\,4686,...) or are only marginally variable (e.g.\ the \od emissions, the unidentified $\lambda\lambda$\,4486, 4504 emissions,...). These various emissions have thus most probably different origins. Several different hypotheses on the wind structure and the nature of \hd\ can be considered:
\begin{enumerate}
\item The emission lines in the spectrum of \hd\ could be formed in jets as suggested by Underhill (\cite{und2}). The long-term variations could be due to a strong occultation of the jets by the stellar body. Such an occultation might result from a precession of the jets. In our data, the absorption continues to strengthen after the emission apparently vanished (e.g.\ \he $\lambda$\,4713) whereas the EWs should reach a limiting value during the eclipse. One possibility could be that in 2000 we are still in a pre-eclipse phase. However several details remain to be addressed. Where do the rather constant emissions (e.g.\ \hee $\lambda$\,4686) originate? What triggers the precession of the jets? Moreover, the narrow emission lines indicate that these jets must be seen almost perpendicular to the line of sight and it seems difficult to imagine a geometric configuration that could explain the secular variations without exhibiting a simultaneous modulation of the emission on the (much shorter) time scale of the stellar rotation period.  
\item The star could be surrounded by a disc which undergoes density oscillations (see e.g.\ the models by Okazaki \cite{oka}). The emission lines would be attenuated when these oscillations are eclipsed by the star, provided that the inclination is sufficient. However, the narrow emission lines without any evidence of a double-peaked profile and the spectropolarimetric observations of Harries (2000) suggest that such a disc should be seen face-on. Therefore it seems difficult to explain the observed variations with this model.  
\item Alternatively, the wind of \hd\ could be confined by a strong enough magnetic field into a high density cooling disc in the plane of the magnetic equator (see Babel \& Montmerle \cite{Babel}). The emission lines would be formed in this cooling disc. If, for some reason, the strength of the magnetic field changes as a function of time, we expect the density in the disc and hence the emission strength to vary. Again, we would have to explain why there is a dichotomy in the  behaviour of the emission lines and what would be the mechanism causing the modulation of the magnetic field strength. 
\item \hd\ might be a binary system harboring a compact object which is revolving in a highly eccentric orbit around an Oe star which is surrounded by a disc seen face-on. Near periastron passage, the X-rays emitted by the accreting compact object (neutron star or black hole) could drastically alter the ionization in the circumstellar enveloppe. The X-rays could completely ionize hydrogen and He~{\sc i}, and the emission in those lines would thus disappear. As long as the compact companion is far enough from the Oe star, the \hee $\lambda$\,4686 formation zone should not be affected too much. If the compact object follows a highly eccentric orbit, the spectral lines should not be affected during most part of the orbit. As some older data are missing (especially from 1955 to 1968), we cannot check what was the morphology of the lines at this epoch. On the other hand, the enhanced interaction around periastron should generate an enhanced X-ray emission. In this context it is worth mentioning that \hd\ was detected as a fairly bright source in the ROSAT All Sky Survey (RASS). In fact, Rutledge et al.\ (\cite{Rut}) quote a ROSAT-PSPC count rate of  $0.0614 \pm 0.0124$\,cts\,s$^{-1}$ for \hd. Assuming that the X-ray emission is produced in an optically thin thermal plasma with $kT = 0.5$\,keV and accounting for an interstellar neutral hydrogen column density of $3.4 \times 10^{21}$\,cm$^{-2}$ (Diplas \& Savage \cite{DS}), we derive an unabsorbed flux of $2.3 \times 10^{-12}$\,erg\,cm$^{-2}$\,s$^{-1}$ in the energy range 0.1 -- 2.0\,keV\footnote{This flux is in very good agreement with the 0.032\,cts\,s$^{-1}$ EINSTEIN-IPC count rate quoted by Chlebowski et al.\ (\cite{Chle}).}. Again assuming a distance of 2.51\,kpc (Gies \cite{gies}), we derive an X-ray luminosity of $L_x = 1.7 \times 10^{33}$\,erg\,s$^{-1}$ corresponding roughly to $\log{L_x/L_{\rm bol}} \sim -6.0$. The ratio between the X-ray and bolometric luminosities in 1990 (i.e.\ at the time of the RASS) was therefore about six times larger than expected from the `canonical' relation proposed by Bergh\"ofer et al.\ (\cite{berg}). Observations of \hd\ with the XMM-Newton satellite might allow to confirm the compact companion hypothesis since they will be obtained when the stars would be close to periastron passage.\\ 
If this compact companion model applies, then the system should certainly have undergone a supernova. Since the runaway status is not established, another confirmation of this scenario could come from the detection of a supernova remnant (SNR) in the vicinity of the star. No such SNR has been detected so far. In the ambient circumstellar medium ionized by the O-star, the SNR signature would be very difficult to see. Another possibility for a null detection could be that this remnant has already evaporated. Maeder \& Meynet (\cite{maed}) predict lifetimes between $2.6 \times 10^6$ and $4.4 \times 10^6$\,yrs for stars of spectral type O3 to O6, whereas a typical O7.5 star stays $4.6 \times 10^6$\,yrs on the main sequence\footnote{Assuming that the typical initial masses of O3, O6 and O7.5 stars are respectively 85, 30 and 20\,M$_{\odot}$.}. Since a SNR takes typically $5 \times 10^4$\,yrs to evaporate, the compact companion hypothesis cannot be ruled out on the grounds of the lack of a detected SNR. 
\end{enumerate}
As the behaviour of \hd\ seems recurrent, and could reproduce after about 56 years, this star certainly deserves a long-term spectroscopic monitoring. The H~{\sc{i}} and \he lines could either reach a steady state or continue their evolution towards stronger absorptions, before they gradually recover a P Cygni profile. Photometric observations of this star should also be useful to assess the amplitude of the possible magnitude variations of \hd, which might be in correlation with its line profile variability. Moreover, multiwavelength studies should be undertaken, in order to confirm possible variations of the mass-loss rate or the compact companion hypothesis. Only then, will we finally be able to answer the question about the real nature of \hd.

\acknowledgement{We would like to thank Dr.\ J.\ Manfroid for his help in collecting additional spectra in 2000 and Drs.\ E.\ Gosset and T.J.\ Harries for discussion. We thank the referee Dr.\ O.\ Stahl for a careful reading of the manuscript and for his valuable suggestions. JMV and GR would like to thank the staff of the Observatoire de Haute-Provence for their technical support during the various observing runs. We are greatly indebted to the Fonds National de la Recherche Scientifique (Belgium) for multiple assistance including the financial support for the rent of the OHP telescope in 1999 and 2000 through contract 1.5.051.00 ``Cr\'edit aux Chercheurs'' FNRS. The travels to OHP for the observing runs were supported by the Minist\`ere de l'Enseignement Sup\'erieur et de la Recherche de la Communaut\'e Fran\c caise. This research is also supported in part by contract P4/05 ``P\^ole d'Attraction Interuniversitaire'' (SSTC-Belgium) and through the PRODEX XMM-OM and Integral Projects. The SIMBAD database has been consulted for the bibliography.}

\end {document}